\documentclass[a4paper]{IEEEtran}

\usepackage{wrapfig,enumerate}

  \usepackage[dvips]{graphicx}
  \graphicspath{{./Figures/}}
  \DeclareGraphicsExtensions{.eps}

\usepackage[cmex10]{amsmath}
\usepackage{amsthm}
\usepackage{amssymb}
\usepackage{color}
\usepackage{comment}

\usepackage{graphicx}
\usepackage{caption}
\usepackage{subcaption}

\usepackage{balance}

\begin{document}

\title{LWIP and Wi-Fi Boost Link Management}
\author{David L\'opez-P\'erez$^1$, Jonathan Ling$^1$, Bong Ho Kim$^1$, Vasudevan Subramanian$^1$, Satish Kanugovi$^1$ Ming Ding$^2$\\
$^1$Bell Laboratories/Nokia\\
$^2$Data 61, Australia}
\maketitle

\IEEEpeerreviewmaketitle

\begin{abstract}

3GPP LWIP Release 13 technology and its pre-standard version Wi-Fi Boost have recently emerged as an efficient LTE and Wi-Fi integration at the IP layer, 
allowing uplink on LTE and downlink on Wi-Fi.
This solves all the contention problems of Wi-Fi and allows an optimum usage of the unlicensed band for downlink.  
In this paper, we present a new feature of Wi-Fi Boost,
its radio link management,
which allows to steer the downlink traffic between both LTE and Wi-Fi upon congestion detection in an intelligent manner. 
This customised congestion detection algorithm is based  on IP probing, and can work with any Wi-Fi access point. 
Simulation results in a typical enterprise scenario show that 
LWIP R13 and Wi-Fi Boost can enhance network performance up to 5x and 6x over LTE-only,
and 4x and 5x over Wi-Fi only networks, respectively,
and that the the proposed radio link management can further improve Wi-Fi Boost performance over LWIP R13 up to 19\,\%.
Based on the promising results, 
this paper suggests to enhance LWIP R13 user feedback in future LTE releases. 
   
\end{abstract}

\section{Introduction}\label{sec:introduction}

Due to the increasing number of more powerful user equipment (UE) and more appealing user applications,
wireless networks have been witnessing and will continue to see an explosive traffic growth in the years to come~\cite{Lopez2015Towards1Gbps}.
Indeed, recent forecasts indicate that 
mobile network operators will need to enhance their network capacity by a factor of 100x in oder to meet their customer demands by 2020~\cite{Report_CISCO}.  
In this context, 
the interworking between Long Term Evolution (LTE)~\cite{Sesia2009} and Wireless Fidelity (Wi-Fi)~\cite{Perahia2013} networks has gained a lot of attention during the last years.
LTE can leverage licensed carriers to realise quality of service and act as a mean of controlling ad-hoc Wi-Fi deployments,
while Wi-Fi itself can allow operators to cost-effectively densify their networks and gain access to a large bandwidth in the unlicensed spectrum. 
The efficient integration of both technologies represents a good opportunity to improve the overall spectral efficiency of future wireless systems and realise effective traffic offloading/aggregation between them both.

In order to realise this efficient LTE and Wi-Fi integration, 
a Third Generation Partnership Project (3GPP) Release~13 standard, 
named LTE Wi-Fi Radio Level Integration with IPsec Tunnel (LWIP), 
is gaining much momentum within the industry~\cite{TS36300}~\cite{TS36331}~\cite{TS36361}.
The foundation for LWIP R13 is Wi-Fi Boost
which realised the first internet protocol (IP) layer LTE and Wi-Fi integration~\cite{Ling2015}~\cite{LingKMV15}~\cite{PIMRCBoost}. 
For LTE and Wi-Fi anchored applications, 
Wi-Fi Boost allows uplink (UL) on cellular and downlink (DL) on Wi-Fi, 
so that UEs can seamlessly and simultaneously draw on the strengths of both networks.  

Wi-Fi is already commonplace in enterprises today, 
but it is not enough and it is not perfect. 
Moreover, Wi-Fi is limited in scalability and quality, and security issues still persist. 
In particular, IT managers are concerned about UL interference problems, poor range and unfair service quality, 
which is granted simply on the proximity of one UE to the access point (AP) compared to another,
the so-called \emph{capture effect}~\cite{1046683}. 
Looking more closely at Wi-Fi's limitations, 
several problems can be traced to the sharing mechanism between the UL and the DL, 
i.e. Wi-Fi's carrier sense multiple access/collision avoidance (CSMA/CA),
as well as the contention between the UE uplinks~\cite{Goldman2011}~\cite{Madan2012}~\cite{Flores2013}. 
In contrast, an LTE-based system does not have this problem of UL conflicts because it uses centralised scheduled access mechanisms. 
 
Wi-Fi Boost, the pre-standard LWIP R13, presents a solution to the above mentioned issues,
and has been firstly targeted at enterprises where it has several benefits:
\begin{itemize}
\item
Wi-Fi Boost uses LTE access for UL and frees up the enterprise's existing Wi-Fi network for DL. 
This means enterprise UEs get the best possible upload and download performance,
as well as excellent indoor cellular coverage through LTE small cells.
\item
Wi-Fi Boost allows operators to leverage vast incumbent Wi-Fi installed APs to supplement LTE capacity. 
The solution works without any hardware or software upgrade on Wi-Fi infrastructure, 
and only requires a software upgrade on LTE small cell BSs and UEs.
\item
A unique feature of Wi-Fi Boost is the local access mode, 
which is a great value added that operators can offer to enterprises beyond just providing an additional access using small cells. 
Local access allows the UE to choose either LTE or Wi-Fi for UL applications anchored in the enterprise core. 
This means much better quality of experience and support of higher capacity for business-impacting enterprise applications such as Lync, Skype, Jabber, Webex, Video conference etc. 
The ability to use Wi-Fi UL when LTE UL degrades ensures better quality of experience for applications anchored in LTE or Wi-Fi core. 
Since Wi-Fi Boost, integrates LTE and Wi-Fi accesses at the IP layer, 
local access is possible via a simple software upgrade involving routing and tunnel configuration on the LTE small cells and UEs.
\end{itemize}

\begin{figure*}
    \centering
    \begin{subfigure}[b]{0.46\textwidth}
    	\centering
        \includegraphics[width=0.85\textwidth]{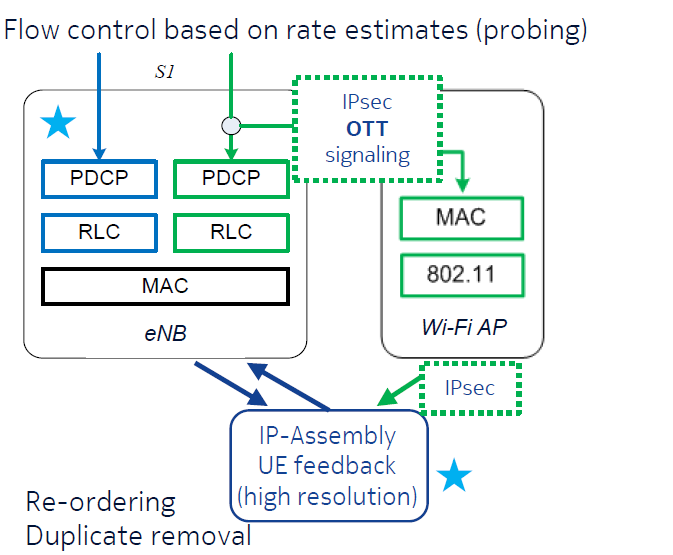}
        \caption{Wi-Fi Boost architecture.}
        \label{fig:boostArch}
    \end{subfigure}
    ~ 
    \vspace{0.1cm}   
    \begin{subfigure}[b]{0.46\textwidth}
    	\centering
        \includegraphics[width=0.85\textwidth]{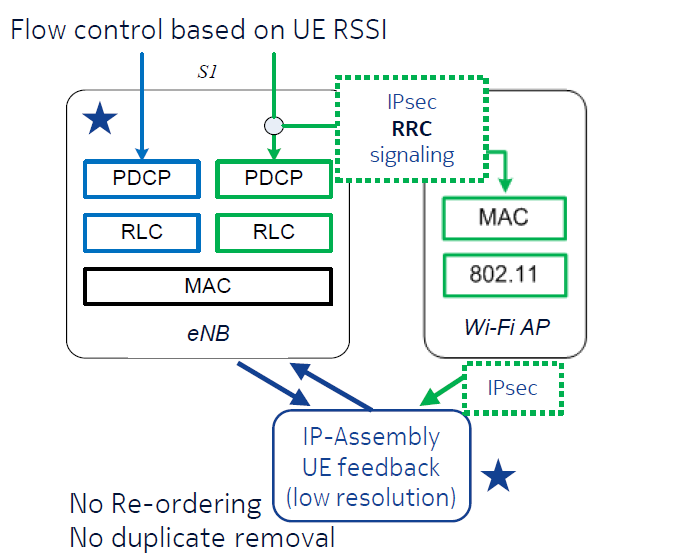}
        \caption{LWIP R13 architecture.}
        \label{fig:lwipArch}
    \end{subfigure}
    \caption{Wi-Fi Boost and LWIP R13 architectures.}\label{fig:architecture}
    \vspace{-0.3cm}
\end{figure*}

Due to these advantages,
Wi-Fi Boost and now its standardised version LWIP R13,
are game changers in the industry:  
\begin{itemize}
\item
They open the door to position small cells into the enterprise market. 
They can also enable new partnerships between LTE operators and Wi-Fi providers, 
especially in environments such as large enterprises and outdoor public Wi-Fi, 
where hundreds of Wi-Fi APs are already deployed and where several additional years of service are expected.
\item
These solutions have been tested for LTE UL but are designed in principle to work on 3G as well. 
With 5G not far away in some markets, 
Wi-Fi Boost and LWIP R13 offers a path towards 4G-5G interworking and in fact multi-technology interworking.
\item
3GPP standardisation of LWIP R13 has been supported by major UE vendors,
and will drive the UE ecosystem. 
The loose integration between LTE and Wi-Fi paths at the IP layer simplifies the device implementation and can potentially be delivered as a software upgrade in existing LTE small cell BSs,
as mentioned before. 
\end{itemize}

Wi-Fi Boost and LWIP R13 technologies are thus the foundation for realising the `all-wireless enterprise' vision,
and represent a significant step forward towards the 5G.

In this paper, we further investigate a new feature of Wi-Fi Boost, 
its radio link management,
which allows to steer the DL traffic between both LTE and Wi-Fi upon congestion detection.
In more detail, it continuously monitors the quality of the Wi-Fi link and moves the UE over to LTE or back to Wi-Fi,
without service interruption. 
The rest of the paper is organised as follows: 
In Section~\ref{sec:boost}, the architectures of Wi-Fi Boost and LWIP R13 are introduced.
In Section~\ref{sec:linkManagement}, the new radio link management devised for Wi-Fi Boost is presented.
In Section~\ref{sec:simulation}, simulation results, which show the performance of Wi-Fi Boost with the proposed radio link management with respect to LTE only, Wi-Fi only and LWIP R13 technologies, are discussed.
Finally, in Section~\ref{sec:conclusions}, the conclusions are drawn.

\section{Boost and LWIP R13 Architectures}\label{sec:boost}

In this section, 
the Wi-Fi Boost and LWIP R13 architectures are presented,
while describing their common features as well as their main differences. 
Fig.~\ref{fig:architecture} illustrates such architectures. 

As mentioned in the introduction,
Wi-Fi Boost is the pre-standards version of LWIP R13,
and thus both technologies share the same interworking philosophy,
as well as other important functionalities.
These main commonalities are described in the following:
\begin{itemize}
\item
Both Wi-Fi Boost and LWIP R13 use as DL anchor the LTE small cell BS,
and utilise as split/aggregation point the IP layer,
as shown in Fig.~\ref{fig:architecture}.
This allows to leverage existing Wi-Fi deployments to supplement LTE capacity without any hardware or software upgrade on the Wi-Fi infrastructure,
which represents a major benefit for operators with vast Wi-Fi rollouts.
Operators can just deploy a reduced number of LTE small cell BSs to control and enhance the performance of an existing large population of Wi-Fi APs.
\item
Both technologies are able to take advantage of the DL and UL split concept 
i.e. UL on LTE and DL on Wi-Fi~\cite{PIMRCBoost}.
By redirecting UL traffic from the Wi-Fi network (unlicensed band) to the LTE network (licensed band).
there is no contention to resolve inside an individual Wi-Fi cell using the CSMA/CA protocol,
which avoids the delay introduced by such contention and ensures a completely collision-free operation inside the cell.
As a result,
Wi-Fi operates only in the DL and works on a cell-centric scheduled basis 
(DL Wi-Fi traffic is scheduled by the Wi-Fi AP),
enabling the most  efficient use of Wi-Fi's large bandwidth,
\item
An IPsec tunnel is used  to transmit DL traffic from the LTE small cell BS to the UE through the Wi-Fi AP in a secure manner. 
It is important to note that the IPsec tunnelling protocol appends an IPsec header to the  DL IP packets that travel from the LTE small cell BS to the UE over the Wi-Fi AP. 
The IPsec overhead is 66\,bytes plus padding
(if the inner IPsec packet plus 2\,bytes IPsec trailer is not a multiple of 16\,bytes, padding is needed). 
Such overhead  may be negligible for large IP packets, 
e.g. file transfer protocol (FTP) packets of 1500\,bytes,
but may be a burden for small IP packets of just few hundreds of bytes. 
However, most of the internet traffic uses IP packets of around 1500\,bytes and this should not be a concern. 
\end{itemize}

Although Wi-Fi Boost and LWIP R13 share the most salient features,
there are differences between them both, 
which are summarised in the following:
\begin{itemize}
\item
One of the main differences between Wi-Fi Boost and LWIP R13 is the way in which the above mentioned IPsec tunnel is set up. 
Due to the lack of a standardised approach,
Wi-Fi Boost uses over the top signalling to establish the IPsec tunnel,
which is a proprietary approach that only requires a software upgrade on LTE small cell BSs. 
In contrast, LWIP R13 benefits from a standardised approach to this,
and the IPsec can be established using layer radio resource control (RCC) signalling. 
\item 
Another important difference between both technologies is the radio link management/traffic steering capabilities at the LTE small cell BS.
Due to its pre-standard and proprietary nature, 
the WiFi-Boost solution allows for a more powerful radio link management,
at the expense of the software upgrades required at the UE side in order to realise the necessary cooperation/feedback.
In contrast, LWIP R13 provides a standardised UE feedback framework,
in which the UE can report RSSI measurements on neighbouring Wi-Fi APs to the LTE small cell BS.
This permits a more universal approach to link management.
Unfortunately, RSSI feedback does not enable congestion detection and thus the traffic steering capabilities are limited
(traffic steering only happens when the strength of the serving path is weak). 
Section~\ref{sec:linkManagement} will present our proposed Wi-Fi Boost radio link management with congestion detection,
and Section~\ref{sec:simulation} will provide a comparison between the performance of such Wi-Fi Boost radio link management with congestion detection
and that of LWIP R13 based on RSSI measurements. 
\item
IP re-ordering and duplicate discard at the UE is another distinctive feature that can be made available in Wi-Fi Boost, 
but it is not present in LWIP R13.
However, it is important to note that since link switching at LWIP R13 only happens when the strength of the serving path is weak,
re-ordering and duplicate discard are not major issues. 
These features become important when considering aggregation, and IP packets arrive to the UE simultaneously via different paths. 
Mobility also makes IP re-ordering and duplicate discard desirable features in the presence of link switching. 
\end{itemize}

\section{Radio Link Management}\label{sec:linkManagement}

In this section, our proposed Wi-Fi Boost radio link management with congestion detection is presented. 
In essence, 
UEs will tend to connect to the Wi-Fi path, 
and switch to the LTE path if congestion is detected in the Wi-Fi path.
UEs may be switched back to the Wi-Fi path if such congestion disappears.
This radio link management could be used as basis for realising load balancing strategies,
but they are out of the scope of this paper. 
As a working assumption, 
we assume that radio link statistics can be obtained from the MAC of the LTE small cell BS for the LTE path,
but that such statistics are not available from the Wi-Fi AP for the Wi-Fi path
(e.g. the Wi-Fi AP may belong to a different manufacturer).
Probing over the Wi-Fi path is used to access its performance and generate the necessary statistics. 

\begin{figure}
	\centering
	\includegraphics[width=0.95\columnwidth]{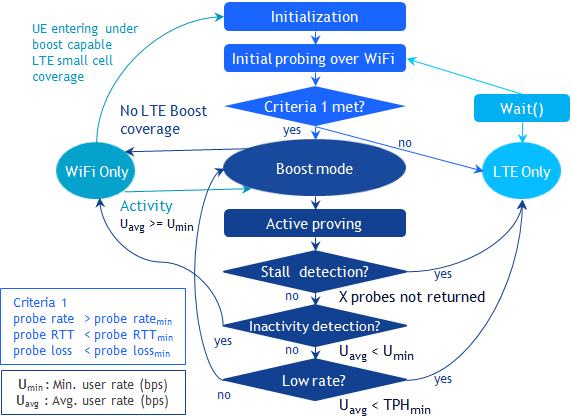}
	\caption{Radio link management diagram.}
	\label{fig:rlm}
	\vspace{-0.3cm}
\end{figure}

In the following, the initial phase (which takes place at connection setup) and data phase (which takes place when UE data is flowing) of the proposed radio link management algorithm are described in detail. 
Fig.~\ref{fig:rlm} illustrates the proposed radio link management algorithm.

\subsection{Initial Phase}

Upon connection request, 
the LTE small cell BS will estimate which is the most suitable path for the given UE. 
This is done by the radio control manager (RCM),
a new logical entity deployed at the LTE small cell BS to realise Wi-Fi Boost.
A UE will be connected to the Wi-Fi path
if the received signal strength of the Wi-Fi pilot is larger than -82\,dBm in the 20\,MHz channel,
and the initial probing results meet a predefined criteria;
Otherwise, it will be connected to the LTE path. 

In order to probe the Wi-Fi path,
the RCM sends to UE $u$ though the Wi-Fi path 
$x^{\rm ini}_u$ IP probes of size $s^{\rm ini}_u$ bits during a test period of $t^{\rm ini}_u$\,s at a rate of $r^{\rm ini}_u$\,Mbps.

Upon probe reception,
the user control manager (UCM),
a new logical entity deployed at the UE to realise Wi-Fi Boost,
gathers the following statistics:
\emph{i)} fraction of probes lost, ${\rm probe\_lost}_u$, 
\emph{ii)} average probe delay, ${\rm probe\_delay}_u$ and 
\emph{iii)} average probe throughput, ${\rm probe\_rate}_u$.
Once the UCM receives the last probe,
this sends back to the RCM a probe ACK with the statistics over such test period. 

Then, if the RCM receives the probe ACK and the following criteria is met:
\begin{itemize}
\item
the fraction of probes lost is lower than a threshold, 
${\rm probe\_lost}_u < {\rm probe\_lost}_{{\rm min},u}$, 
\item
the average probe delay is shorter than a threshold, 
${\rm probe\_delay}_u < {\rm probe\_delay}_{{\rm min},u}$, and 
\item
the average probe rate is higher than a threshold, 
${\rm probe\_rate}_u < {\rm probe\_rate}_{{\rm min},u}$,
\end{itemize}
the RCM routes UE IP data packets over the Wi-Fi path;
Otherwise, it will route packets over the LTE path. 
Note that all mentioned thresholds are quality of service dependent. 

It is important to note that if the UE is routed over the LTE path,
the RCM will proceed with an initial probing phase every $t^{\rm ip}_u = 2$\,s during the data transmission
in order to check wether the Wi-Fi path is suitable for the transmission. 

\subsection{Data Phase}

If the Wi-Fi path has been selected,
the RCM performs active probing to check whether the quality of the Wi-Fi path is still suitable to carry UE's traffic,
or it has degraded due to congestion.
In the latter case, the RCM would switch the UE over the LTE path. 

In order to probe the Wi-Fi path while the actual data transmission is taking place,
the RCM sends to the UE though the Wi-Fi path 
an active IP probe of size $s^{\rm dat}_u$ bits inserted within the UE IP data packets every $t^{\rm dat}_u$ seconds.

Upon active probe reception, 
the UCM calculates the average UE throughput, $U_{{\rm avg}, u}$, in between this active probe and the previous one,
and feeds back to the RCM the computed value using an active probe  ACK.
In contrast to the initial probes where only the last probe was acknowledged, 
all active probes are acknowledged.

Then, upon active probe ACK reception,
the RCM puts the average UE throughput, $U_{{\rm avg}, u}$, over a moving average filter, ${\hat U}_{{\rm avg}, u}$,
and may take the following decisions:
\begin{itemize}
\item
Stall detection: 
If $x_{\rm stall}$ consecutive active probe ACKs are missing,
the RCM switches the UE over the LTE path.
\item
Inactivity detection:
If the number of bits transmitted in between two probe ACKs is smaller than a threshold,
$U_{{\rm avg}, u} < U_{{\rm min}, u}$,
meaning that the UE generates a small amount of traffic,
the RCM switches the UE to Wi-Fi only mode in order to save Wi-Fi Boost resources. 
\item
Congestion detection: 
if the filtered average UE throughput is smaller than a threshold,
${\hat U}_{{\rm avg}, u} < TPH_{{\rm min}, u}$,
the RCM switches the UE over the LTE path.
\end{itemize}
Otherwise, the RCM keeps the UE IP data packets over the Wi-Fi path.
Note that all mentioned thresholds are quality of service dependent. 

It is important to note that if the UE is re-routed over the LTE path,
the RCM will proceed with an initial probing phase every $t_{\rm ip}$ seconds
in order to check wether the Wi-Fi path is suitable for the transmission. 

Moreover, the following constraints to switching apply:
\begin{itemize}
\item
No more than one UE is switched very $t_{\rm switch}$ seconds in any direction
in order to avoid massive switching and instability issues in the presence of congestion.
\item
No UE is switched to the LTE path, 
if the resulting average UE throughputs of the existing LTE UEs after switching such UE would be smaller than a threshold $r_{\rm switch}$.
Round robin assumptions can be used to estimate the LTE UE performance after switching.  
\end{itemize}

\section{Simulation Results}\label{sec:simulation}

In this section, simulation results are presented to validate the performance of presented Wi-Fi Boost radio link management in terms of FTP capacity.
The performance evaluation is conducted over an enterprise scenario of 50\,m$\times$120\,m,
where there is a LTE small cell {BS} located at the centre of it and several {Wi-Fi} {APs} are deployed within the enterprise. 
Most simulation assumptions
in terms of {BS} and {UE} deployment as well as antenna gain, path loss, shadowing and multi-path fading modelling
follow the {3GPP} recommendations in~\cite{TR36889}.
Since the focus is only on the DL performance,
UL performance is not characterised in the paper. 
The assumption is that there is enough {UL} bandwidth to accommodate the UL diverted traffic, e.g. TCP ACKs, data channels.
100 simulation drops are performed, and in each drop 10 seconds are simulated. 
Please refer to~\cite{PIMRCBoost} for a more complete description of the simulator. 


\paragraph{{Wi-Fi} {AP} deployment} 
2 {Wi-Fi} channels of 20\,MHz in the 5\,GHz band are considered,
and 2 {AP} are deployed in the enterprise where the inter-{AP} distance is 60\,m.
Each {AP} has a transmit power of 24\,dBm,
and selects upon deployment the channel in which the least load and interference is observed. 
Two omnidirectional antennas with a 5\,dBi gain are considered.

\paragraph{{UE} deployment}
1, 4, 20, 26 or 32 {UE} are uniformly deployed within the enterprise,
where the minimum {AP}-to-{UE} distance is 3\,m.
Each {UE} has a transmit power of 18\,dBm, 
and associates to the {AP} with the strongest pilot,
provided that this pilot was detected at or above $-$82\,dBm in the 20\,MHz channel.
Two omnidirectional antennas with a 0\,dBi gain are considered,
thus allowing 2$\times$2 MIMO transmissions. 
Fast fading channel gains are driven a UE speed of 3km/h.

\paragraph{Services}
All {UEs} use a bidirectional {FTP} service (3GPP FTP traffic model~2)
The {FTP} file size is 0.5\,Mbytes  in the DL and 0.25\,Mbytes in the UL, 
while the mean reading time is 0.1\,s 
(leading to a high demand of 40\,Mbps and 20\,Mbps per UE in DL and UL respectively).
Note that {TCP} {ACK} are generated in response to {FTP} traffic,
where 1~{TCP} {ACK} is sent for every 3~{TCP} data packets.

Other relevant Wi-Fi parameters are set as follows: 
DIFS$=34\,\mu$s, SIFS$=16\,\mu$s, time slot $= 9\,\mu$s, TXOP$= $3\,ms.

\subsection{Benchmarked Technologies}

Four system configurations are  considered:
\begin{enumerate}
\item 
\emph{LTE only}: 
All traffic DL and UL is carried by the LTE small cell BS in the licensed band.
\item 
\emph{Wi-Fi only}: 
All traffic DL and UL is carried by the Wi-Fi APs in the unlicensed band.
\item 
\emph{LWIP R13}: 
Traffic is split according to the discussion in Section~\ref{sec:boost}.
DL FTP traffic and DL TCP ACKs are routed over WiFi, 
while UL FTP traffic and UL TCP ACKs are routed over LTE.  
As explained before, WiFi MAC ACKs remain in the WiFi network.  
Note that due to the static nature of the UEs in our simulation, 
LWIP R13 cannot leverage its RSSI radio link management. 
\item 
\emph{Wi-Fi Boost}:
Traffic is split as in the LWIP R13 case.
However, the congestion detection and DL steering  mechanism presented in Section~\ref{sec:linkManagement} kicks in to optimise overall enterprise performance
when the UE does not get the desired performance.
\end{enumerate}

Note that the Wi-Fi Boost radio link management is configured with the following parameters:
\begin{itemize}
\item
Initial phase probing:\\
$s^{\rm ini}_u = 12000$\,bits,
$t^{\rm ini}_u = 0.1$\,s and
$r^{\rm ini}_u = 5$\,Mbps.
\item
Initial phase decision-making:\\
${\rm probe\_lost}_{{\rm min},u} = 0.9$, 
${\rm probe\_delay}_{{\rm min},u} = 0.5$\,s and 
${\rm probe\_rate}_{{\rm min},u} = 5$\,Mbps.
\item
Data phase probing:\\
$s^{\rm dat}_u = 160$\,bits and
$t^{\rm dat}_u = 0.003$\,s.
\item
Data phase decision-making:\\
$x_{\rm stall} = 3$,
$U_{{\rm min}, u} = 0.5$\,Mbps and
$TPH_{{\rm min}, u} = 5$\,Mbps.
\end{itemize}
The IPsec overhead is 66\,bytes.

\subsection{Performance Comparison}

For reference purposes and according to our simulations,
let us first note that the peak UE throughput (single UE case) for the LTE only case was 63\,Mbps,
while that for the Wi-Fi only case was 140\,Mbps. 
As LWIP R13 and Wi-Fi Boost do not leverage aggregation,
their peak UE throughput were equal to that of the Wi-Fi only case, 135\,Mbps.

\begin{figure}
	\centering
	\includegraphics[width=1\columnwidth]{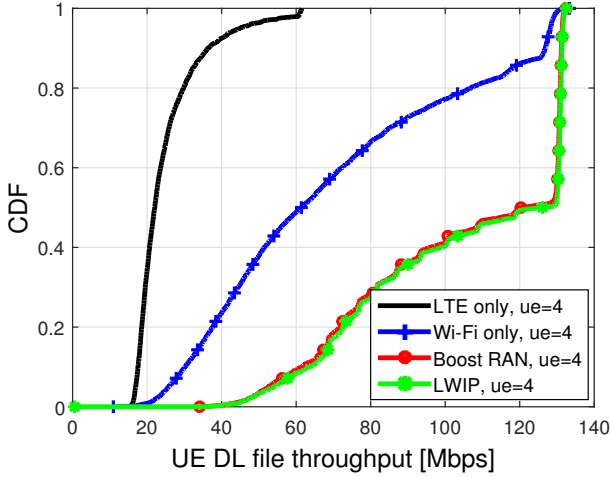}
	\caption{UE throughput distribution for the 4 user per enterprise case.}
	\label{fig:4ueTP}
		\vspace{-0.3cm}
\end{figure}

Fig.~\ref{fig:4ueTP} shows the UE throughput distribution for the case where there are 4 UEs in the enterprise.
The LTE only case provides a median throughput of 21.93\,Mbps/UE,
while the Wi-Fi only case provides a larger median throughput of 60.93\,Mbps/UE.
This is because the Wi-Fi only case benefits from more cells (2 instead of 1), more bandwidth (2x20MHz instead of 1x10MHz) and a larger peak modulation (256QAM instead of 64QAM).
Results also show that LWIP R13 and Wi-Fi Boost have a substantial gain over the Wi-Fi only case of around 2x.
This is due to the offloading of UL traffic from the unlicensed to the licensed band and the resulting collision-free usage of the unlicensed spectrum for DL (the so-called Boost effect).
Note that LWIP R13 and Wi-Fi Boost perform equally.
Because of the low load in the scenario,
there is no congestion and the radio link management of Wi-Fi Boost presented in this paper does not kick in.

\begin{figure}
	\centering
	\includegraphics[width=1\columnwidth]{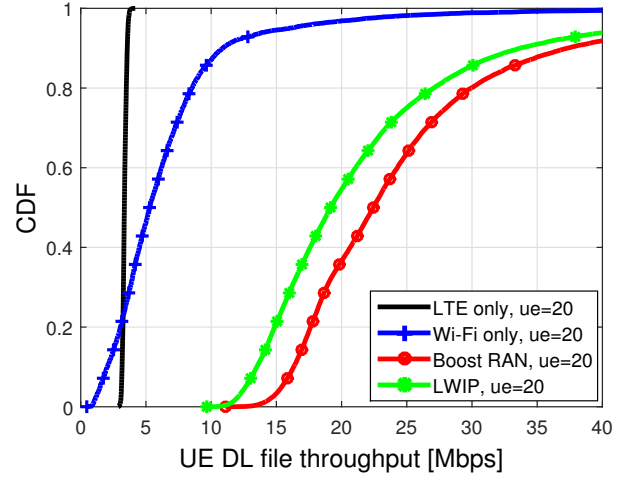}
	\caption{UE throughput distribution for the 20 user per enterprise case.}
	\label{fig:20ueTP}
		\vspace{-0.3cm}
\end{figure}

Fig.~\ref{fig:20ueTP} shows the UE throughput distribution for the case where there are 20 UEs in the enterprise.
Now, the LTE only case provides a median throughput of 3.37\,Mbps/UE,
while the Wi-Fi only case provides just a slightly larger median throughput of 7.3\,Mbps/UE.
Even if Wi-Fi has more cells, more bandwidth and and a larger peak modulation,
the Wi-Fi performance is significantly degraded in comparison to that of LTE 
due to the inefficient sharing of resources between nodes and the contention/collision issues in the former.
Moreover, 
and as in the previous scenario,
LWIP R13 significantly outperforms the Wi-Fi only case with a gain of around 2.6x due to the Boost effect.
The gain is larger than before because the larger load and contention degrades further the performance of the benchmark, 
the performance of Wi-Fi.
It is important to note that this time
Wi-Fi Boost provides a 17\,\% gain over LWIP R13.
Because of the larger load in the scenario, 
the congestion detection mechanism is activated and some UEs are switched from Wi-Fi to LTE,
thus providing a better sharing of overall resources with the subsequent increase in performance. 

\begin{figure}
	\centering
	\includegraphics[width=1\columnwidth]{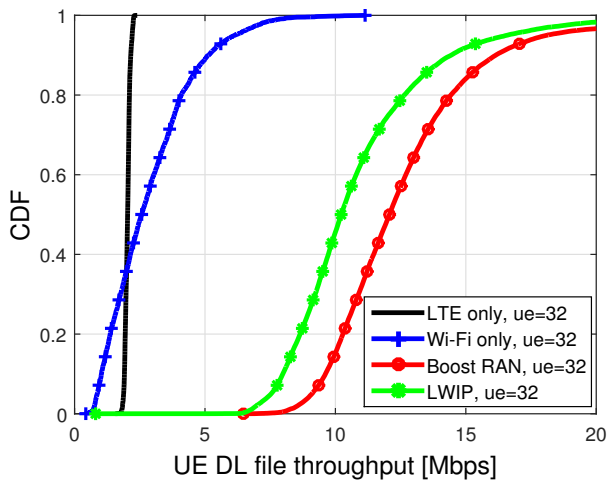}
	\caption{UE throughput distribution for the 32 user per enterprise case.}
	\label{fig:32ueTP}
	\vspace{-0.3cm}
\end{figure}

Fig.~\ref{fig:32ueTP} shows the UE throughput distribution for the case where there are 32 UEs in the enterprise.
Due to the larger load, and the resulting larger contention and congestion,
the gap between the performance of the LTE only and Wi-Fi only cases reduces further.
This shows how CSMA/CA becomes more and more inefficient as the traffic load increases.
Moreover, due to the larger congestion,
the performance gain  of LWIP R13 and Wi-Fi Boost with respect to the Wi-Fi only case is again larger.
LWIP R13 and Wi-Fi Boost can enhance network performance up to 5x and 6x over LTE only,
and 4x and 5x over Wi-Fi only networks, respectively,
which is in inline with the results in~\cite{PIMRCBoost}.
For the same reason, due to the larger congestion,
the  performance gain of LWIP R13 over Wi-Fi Boost is also larger,
around 19\,\%. 
This shows how an intelligent selection of the serving path that does not only relay on RSSI measurements 
can provide a better LTE and Wi-Fi interworking and enhance the UE performance.
This indicates the need for enhancing  LWIP R13 UE feedback in future LTE releases 
by providing UE estimations to the LTE small cell BS on short-term throughput to detect congestion.

\begin{figure}
	\centering
	\includegraphics[width=1\columnwidth]{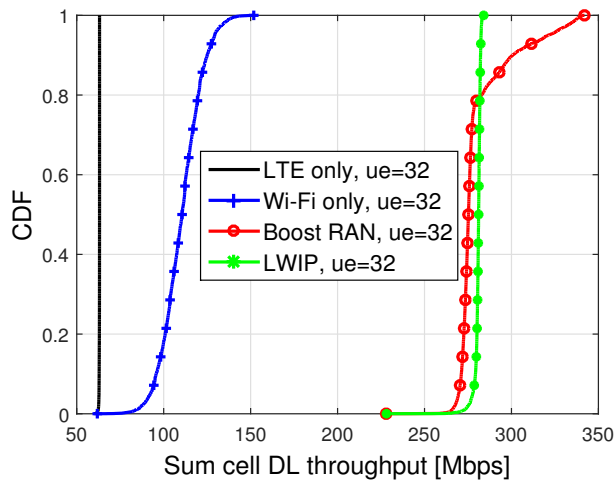}
	\caption{System throughput (sum throughput of all cells) distribution for the 32 user per enterprise case.}
	\label{fig:sector32ueTP}
	\vspace{-0.3cm}
\end{figure}

Fig.~\ref{fig:sector32ueTP} shows the sum cell throughput distribution for the case where there are 32 UEs in the enterprise.
The sum cell throughput is the sum of throughput of all cells in the scenario.
Since there is only one LTE small cell BS in the scenario,  
the LTE only case provides a median cell sum throughput of 63\,Mbps, around its peak throughput. 
Instead, 
the Wi-Fi only case provides a median cell sum throughput of 110\,Mbps.
Congestion prevents achieving the peak throughput of the Wi-Fi cells.
For the LWIP R13 case, 
since such contention disappears due to the Boost effect,
the system reaches the Wi-Fi peak throughput, i.e., 2$\times$140\,Mbps $=$ 280\,Mbps.
Finally, results show how around 20\,\% of the time congestion is detected,
the proposed radio link management mechanism kicks in,
and some UEs are switched to the LTE small cell BS.
This allows to leverage the licensed spectrum achieving a top throughput of up to 340\,Mbps,
around the combined peak throughput of all cells together. 
A more aggressive switching with a larger ${\rm probe\_rate}_{{\rm min},u}$ would provide a better use of the licensed spectrum.
However, this may come at the expense of underutilising the unlicensed spectrum,
which results in a overall degradation of UE and sum throughput.
We recommend to explore machine learning techniques to optimise the parameters of the algorithm with respect to the scenario conditions. 

\vspace{0.25cm}
\section{Conclusion}\label{sec:conclusions}

In this paper, we have presented the architectures of 3GPP LWIP R13 technology and its pre-standard version Wi-Fi Boost,
while highlighting its main common features and differences.
Moreover, we have also proposed a Wi-Fi Boost radio link management with congestion detection to make LTE and Wi-Fi integration more efficient,
where such congestion detection mechanism is based on IP probing and can work with any Wi-Fi AP.
In essence, 
UEs will tend to connect to Wi-Fi path, 
and switch to the LTE path if congestion is detected in the Wi-Fi path.
UEs may be switched back to the Wi-Fi path if such congestion disappears.
Simulation results in a typical enterprise scenario show that 
LWIP R13 and Wi-Fi Boost can enhance network performance up to 5x and 6x over LTE-only,
and 4x and 5x over Wi-Fi only networks, respectively,
and that the the proposed radio link management can further enhance Wi-Fi Boost performance over LWIP R13 up to 19\,\%.
Based on lessons learned, 
this paper suggests to enhance LWIP R13 UE feedback in future LTE releases 
by providing UE estimations to the LTE small cell BS on short-term throughput to detect congestion. 
This would allow to realise the presented radio link management for Wi-Fi Boost in LWIP. 

\bibliographystyle{IEEEtran}

\bibliography{references}

\end{document}